\documentclass[
  twocolumn,
  prl,
  amsmath,
  amssymb,
  superscriptaddress,
  floatfix
]{revtex4}

\usepackage{bm}
\usepackage{graphicx}

\begin{document}

\title{Theory of Anomalous Quantum Hall Effects in Graphene}

\author{P.~M.~Ostrovsky}
\affiliation{
 Institut f\"ur Nanotechnologie, Forschungszentrum Karlsruhe,
 76021 Karlsruhe, Germany
}
\affiliation{
 L.~D.~Landau Institute for Theoretical Physics RAS,
 119334 Moscow, Russia
}

\author{I.~V.~Gornyi}
\altaffiliation{
 Also at A.F.~Ioffe Physico-Technical Institute,
 194021 St.~Petersburg, Russia.
}
\affiliation{
 Institut f\"ur Nanotechnologie, Forschungszentrum Karlsruhe,
 76021 Karlsruhe, Germany
}

\author{A.~D.~Mirlin}
\altaffiliation{
 Also at Petersburg Nuclear Physics Institute,
 188350 St.~Petersburg, Russia.
}
\affiliation{
 Institut f\"ur Nanotechnologie, Forschungszentrum Karlsruhe,
 76021 Karlsruhe, Germany
}
\affiliation{
 Inst. f\"ur Theorie der kondensierten Materie,
 Universit\"at Karlsruhe, 76128 Karlsruhe, Germany
}

\maketitle

\textbf{
Recent successes in manufacturing of atomically thin graphite samples
\cite{Novoselov04} (graphene) have stimulated intense experimental and
theoretical activity \cite{Geim07, RMP07}. The key feature of graphene is the
massless Dirac type of low-energy electron excitations. This gives rise to a
number of unusual physical properties of this system distinguishing it from
conventional two-dimensional metals. One of the most remarkable properties of
graphene is the anomalous quantum Hall effect \cite{Novoselov05, Novoselov07,
Zhang, Jiang, Abanin07}. It is extremely sensitive to the structure of the
system; in particular, it clearly distinguishes single- and double-layer
samples. In spite of the impressive experimental progress, the theory of quantum
Hall effect in graphene has not been established. This theory is a subject of
the present paper. We demonstrate that the Landau level structure by itself is
not sufficient to determine the form of the quantum Hall effect. The Hall
quantization is due to Anderson localization which, in graphene, is very
peculiar and depends strongly on the character of disorder \cite{OurEPJ}. It is
only a special symmetry of disorder that may give rise to anomalous quantum Hall
effects in graphene. We analyze the symmetries of disordered single- and
double-layer graphene in magnetic field and identify the conditions for
anomalous Hall quantization.
}

As was discovered~\cite{vonKlitzing80} in 1980, the Hall conductivity
$\sigma_{xy}$ of a 2D electron gas in a strong transverse magnetic field
develops plateaus at values quantized in units of $e^2/h$. This phenomenon is
the famous integer quantum Hall effect (QHE) \cite{QHEbook} --- one of the most
fascinating quantum effects in the condensed matter physics.

The experimentally measured Hall conductivity of single-layer graphene
\cite{Novoselov05, Novoselov07, Zhang} is quantized taking the odd multiples of
the quantum $2e^2/h$ (here the factor of two is due to the spin degeneracy)
\begin{equation}
 \sigma_{xy}
  = (2k + 1) 2e^2/h, \qquad k\in\mathbb{Z}.
 \label{odd}
\end{equation}
In double-layer samples, the quantum Hall plateaus occur at even multiples of
$2e^2/h$ excluding $k=0$. Owing to this unusual quantization, the Hall
measurements are widely used in modern experiments for characterizing the
graphene samples.  Remarkably, the signatures of Hall conductivity quantization
in graphene were recently observed even at room temperature \cite{Novoselov07}.

A simple argument in favour of the odd QHE (\ref{odd}) in a single graphene
layer \cite{GusyninSharapov05} is based on the structure of Landau levels for
two-dimensional massless electrons \cite{ZhengAndo02}. In clean graphene, the
energies of Landau levels are $\epsilon_N = \hbar \omega_c \mathop{\mathrm{sgn}}
N \sqrt{|N|}$ with $\omega_c = v_0 \sqrt{2 e B/\hbar c}$ and $N\in \mathbb{Z}$.
The plateaus of the Hall conductivity are then identified with its classical
values $\sigma_{xy} = n_e e c/B$ at concentrations $n_e$ corresponding to an
integer filling factor $n = n_e hc/eB$, that is, to an integer number of filled
Landau levels. This consideration is further extended by the calculation of the
Hall conductivity in the presence of disorder within Boltzmann
\cite{GusyninSharapov05} or self-consistent Born approximation
\cite{PeresGuineaCastroNeto06}, i.e., for disorder-broadened Landau levels.

However, the spectral gaps in the density of states between separated Landau
levels do not lead to the QHE. Indeed, while the dependence of
the Hall conductivity on the Fermi energy is quantized in the clean system,
this is not the true QHE. The point is that the Fermi level
itself is not a smooth function of the density: it jumps between the fully
occupied and empty Landau levels with increasing density. As a result, the
density dependence of $\sigma_{xy}$ (which is measured in experiments) shows up
no steps and no plateaus, i.e. no QHE. Including the Landau
level broadening by disorder leads only to the magneto-oscillations of
$\sigma_{xy}(n_e)$ but not to its quantization \cite{Ando}. In fact, (i) the QHE
does not require quantization of the density of states at all; (ii) the position
of the QHE plateaus transition does not necessarily correspond to the center of
Landau level; (iii) the crucial ingredient responsible for the Hall quantization
is the disorder-induced Anderson localization (for review, see
\cite{QHEbook,Huckestein95,EMRMP}).

The existence of the odd-integer (\ref{odd}) QHE in graphene also requires a
more rigorous justification in view of quantum interference effects that are
essential in any two dimensional system, including graphene. Once disorder is
fully taken into account, the quantization of Hall conductivity is exact and the
transition between quantum Hall plateaus becomes a quantum phase transition with
universal critical properties. This immediately shows the non-universality of
the result (\ref{odd}). Indeed, if the disorder in graphene is of a generic form
and does not possess any special symmetry, then the Dirac nature of excitations
will be completely lost at large length scales. This is exactly what occurs in
graphene with a generic (preserving only the global time-reversal symmetry) 
disorder at zero magnetic field $B$. In such a system, localization yields
vanishing conductivity with lowering temperature \cite{Aleiner06, Altland06}.
The critical properties of the generically disordered graphene will not differ
from those for any other two-dimensional system. The quantized Hall conductivity
will then take all integer multiples of $2e^2/h$ rather than the odd series Eq.\
(\ref{odd}). Furthermore, $\sigma_{xy} = 0$ at the Dirac point due to the
particle-hole symmetry. The conventional theory of the QHE \cite{Pruisken84}
predicts complete localization ($\sigma_{xx}=0$) and the plateau in the Hall
conductivity instead of the quantum Hall transition in this situation.
Therefore, the observation of the odd quantization (\ref{odd}) is a striking
experimental result calling for theoretical explanation.

The only reason for a non-standard quantization of the Hall conductivity is the
presence of some special symmetry that is preserved by disorder and thus changes
the critical behavior of the system. Unconventional transport and localization
properties of graphene with special symmetries of disorder at $B=0$ were studied
in Refs. \cite{OurPRB, OurPRL} (see also earlier works on disordered Dirac
fermions \cite{DisorderedDiracFermions, Bocquet00Altland02}). However, the
quantum localization effects (and hence most of the peculiarities arising from
the symmetry of disorder) were discarded in most analytical studies devoted to
the QHE in graphene \cite{GusyninSharapov05, PeresGuineaCastroNeto06, Herbut07,
Gorbar07}. Recent numerical simulations of disordered graphene in magnetic field
\cite{Koshino07, Sheng06, Chakravarty07} have indeed shown that the result is
sensitive to symmetry properties of disorder.

In this paper we develop the theory of the integer QHE in graphene. We carry out
the symmetry analysis and identify the situations when the QHE is anomalous.

\textbf{\textit{Graphene: model and symmetries.}}---
We start with the effective Hamiltonian for the clean single-layer graphene in
external magnetic field
\begin{equation}
 H
  = v_0 \tau_3 \bm{\sigma}\left(
      \mathbf{p} + \frac{e}{c} \mathbf{A}
    \right).
 \label{ham}
\end{equation}
Here the Pauli matrices $\sigma_i$ and $\tau_i$ operate in the space of two
sublattices $A$ and $B$ and two valleys, $K$ and $K'$, of the graphene spectrum,
respectively. The full symmetry classification for this Hamiltonian in the
absence of magnetic field was developed in Ref.\ \cite{OurPRB}. When the
magnetic field is applied, time-inversion symmetry is broken and we are left
with (i) an SU(2) isospin symmetry in the space of valleys \cite{McCann06},
generated by $\Lambda_{x,y} = \sigma_3 \tau_{1,2}$ and $\Lambda_z = \sigma_0
\tau_3$ and (ii) an additional discrete chiral symmetry $C_0$ that arises
exactly at zero energy: $H = - \sigma_3 H \sigma_3$. Further, we denote
$C_{x,y,z}$ the combinations of $C_0$ transformation with the isospin
rotations.

We first consider the situation when all chiral symmetries are broken. This
always happens when the Fermi energy is shifted away from the Dirac point by the
gate voltage. At zero Fermi energy, the chiral symmetry can be violated by e.g.
any potential disorder. In this case we have only two possibilities with respect
to the symmetry: decoupled ($\Lambda_z$ preserved) or mixed ($\Lambda_z$
violated) valleys.

\textbf{\textit{Decoupled valleys: odd quantum Hall effect.}}---
We start with considering the case of decoupled valleys. A physical realization
is any disorder smooth on the scale of lattice spacing. The isospin of electrons
(valley index) is hence preserved. We will show that it is the isospin symmetry
that is responsible for the odd quantization Eq.\ (\ref{odd}). Indeed the
isospin degeneracy implies the quantization of Hall conductivity with the step
$4e^2/h$ (the factor $4$ accounts for $2$ degenerate spin states and $2$
independent valleys). Then to prove the validity of Eq.\ (\ref{odd}), it
suffices to establish the quantum Hall transition at zero filling. In order to
do this, we make use of the low-energy theory (non-linear sigma model
\cite{Efetov}) for disordered graphene with decoupled valleys derived in Ref.\
\cite{OurPRL} (see also Ref.\ \cite{Mudry07}). The model is separated into two
independent sectors corresponding to the two valleys. In each sector the action
has the form
\begin{equation}
 S[Q]
  = \frac{1}{4} \mathop{\mathrm{Str}} \left[
       -\frac{g_{xx}}{2} (\nabla Q)^2
       +\biggl( g_{xy} \pm \frac{1}{2} \biggr)
         Q \nabla_x Q \nabla_y Q
    \right].
 \label{action}
\end{equation}
The field $Q$ is the $4 \times 4$ supermatrix operating in Fermi-Bose and
advanced-retarded spaces. The two parameters of the model, $g_{xx}$ and
$g_{xy}$, are longitudinal and Hall conductivities per one valley and per spin
component measured in units $e^2/h$. The `Str' operation implies the supertrace
in all indices of the matrix along with real-space integration. This action
differs from the usual sigma model in quantizing magnetic field
\cite{Pruisken84} by the addition of $\pm 1/2$ to $g_{xy}$. This additional
contribution arises due to the quantum anomaly of Dirac fermions
\cite{Haldane88, Bocquet00Altland02, OurPRL}. It is the only reminiscent of the
Dirac nature of excitations that survives at large scales and influences the
critical properties. The signs in front of the anomalous terms $1/2$ are
opposite for the two valleys. This ensures the global parity symmetry ($x
\mapsto -x$, $K \mapsto K'$) of the total action.

The second term of the action (\ref{action}) has a topological nature:
$\mathop{\mathrm{Str}} ( Q \nabla_x Q \nabla_y Q) \equiv 8 i \pi N[Q]$ with
$N[Q]$ taking only integer values. This term gives the imaginary part of the
action $\mathop{\mathrm{Im}} S[Q] = \theta N[Q]$ with the vacuum angle
$\theta = 2 \pi g_{xy} \pm \pi$.

\begin{figure}
\centerline{\includegraphics[width=0.9\columnwidth]{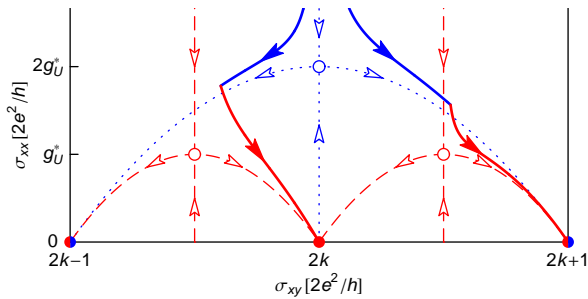}}
\vspace*{-0.2cm}
\caption{Renormalization group flow of $\sigma_{xx}$ and $\sigma_{xy}$ in
graphene with decoupled and mixed valleys. Dotted/dashed lines are separatrices
of the flow for graphene with decoupled/mixed valleys. Open circles are unstable
fixed points corresponding to quantum Hall transitions. Stable fixed points
(plateaus) are shown as disks. Two solid curves demonstrate a possible flow
towards even- and odd-plateau fixed point for a model with weakly mixed valleys.
Each curve has a cusp when the running scale reaches $l_\text{mix}$.}
\label{fig:flow}
\vspace*{-0.5cm}
\end{figure}

The initial values of $g_{xx}$ and $g_{xy}$ are determined by the corresponding
Drude expressions (see Supplementary Materials). The quantum corrections that
establish localization, and hence the QHE, are the result of
renormalization of the action Eq.\ (\ref{action}). The renormalization flow of
$g_{xx}$ and $\theta$ was proposed in Ref.\ \cite{Pruisken84, Khmelnitskii84}.
We plot this flow schematically in Fig.\ \ref{fig:flow} by dotted lines. The
effective theory (\ref{action}) is invariant with respect to the vacuum angle
shift $\theta \mapsto \theta + 2\pi$ hence the flow pattern is periodic function
of $g_{xy}$. Transitions between quantum Hall plateaus occur when the value of
$\theta$ passes through an odd multiple of $\pi$. Owing to the anomalous
contribution in Eq.\ (\ref{action}), this is the case at zero filling factor
when $g_{xy} = 0$. Thus we have shown the validity of the odd quantization
series Eq.\ (\ref{odd}) in the case when \emph{disorder does not mix the
valleys}. The absence of anomaly would have led to a plateau rather than the
transition at $n=0$ similarly to ordinary QHE.

Physically, the step of Hall conductivity between plateaus is due to a critical
delocalized state which is exactly at the Fermi energy when $\theta = \pi$. All
other states are localized and do not contribute to either longitudinal or Hall
conductivity. The value of longitudinal conductivity exhibits a peak at the
transition point with the maximum value 
\begin{equation}
 \sigma_{xx}
  = 4\times g^*_U\approx 2 e^2/h,
 \label{sxx}
\end{equation}
where $g^*_U$ is the longitudinal conductivity for the ordinary quantum Hall
effect (known to be in the range $g^*_U \simeq 0.5 \div 0.6$ from numerical
simulations \cite{Quantum-Hall_critical}) and the factor $4$ again reflects the
valley and spin degeneracy. Equation (\ref{sxx}) agrees with the experimental
value found in strong magnetic field at the Dirac point \cite{Novoselov05,
Zhang, Abanin07}.

\textbf{\textit{Valley mixing: ordinary quantum Hall effect.}}---
Let us now turn to the case when a weak valley mixing is present. For instance,
charged impurities scatter electrons between valleys at some small rate
$\tau_\text{mix}^{-1}$ as compared to the intra-valley scattering rate
$\tau^{-1}$. The total action of the system will then be perturbed by the small
coupling between matrices $Q_K$ and $Q_{K'}$ corresponding to the two valleys
\begin{equation}
 S[Q_K, Q_{K'}]
  = S[Q_K] + S[Q_{K'}]
    + \frac{\hbar \rho}{\tau_\text{mix}} \mathop{\mathrm{Str}} Q_K Q_{K'},
 \label{S12}
\end{equation}
where $S[Q_{K,K'}]$ is given by Eq.\ (\ref{action}), and $\rho$ is the density
of states at the Fermi level (see Supplementary Materials for the derivation).
This perturbation is relevant and leads to the constraint $Q_K = Q_{K'}$ in the
infrared limit. The corresponding valley-mixing length is determined by the
relation $l_\text{mix}/l \sim (\tau_\text{mix} / \tau)^{1/z}$. The ultraviolet
scales $l$ and $\tau$ are given by the effective mean-free path and time; in
strong magnetic field (for low-lying Landau levels with $|N|\sim1$), the length
$l$ is of the order of the magnetic length: $l \sim l_B = \sqrt{\hbar c/e B}$
and the mean free time $\tau$. The index is $z=2$ for non-interacting electrons
(diffusion propagation) and in the case of short-range interaction
\cite{Wang00}. A different value $z$ emerges in the case of Coulomb interaction
\cite{PruiskenBurmistrov}; experiments \cite{Engel93} yield $z \simeq 1$.

At a scale larger than $l_\text{mix}$, we have $Q_K = Q_{K'}$ and the
topological terms with anomalous factors $\pm 1/2$ cancel in Eq.\ (\ref{S12}).
We end up with the unitary sigma model for the \emph{normal} QHE
\cite{Altland06} with $\theta = 4\pi g_{xy}$ and ordinary quantization of Hall
conductivity
\begin{equation}
 \sigma_{xy}
  = k\: 2e^2/h, \qquad k\in\mathbb{Z}.
\label{ordinary}
\end{equation}

\begin{figure}
\centerline{\includegraphics[width=0.9\columnwidth]{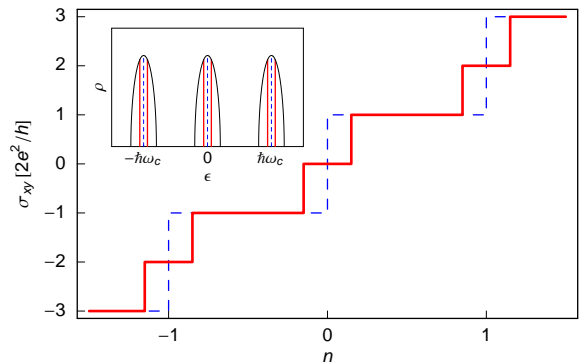}}
\vspace*{-0.2cm}
\caption{Quantum Hall effect in graphene with smooth disorder at zero
temperature. Hall conductivity as a function of the filling factor: odd
(decoupled valleys, dashed line) vs normal (weak valley mixing, solidline)
quantization. Inset shows the energy dependence of the density of states. The
state in the center of Landau level is delocalized (dashed lines) when the
valleys are decoupled. The valley mixing splits this delocalized state (solid
lines).}
\label{fig:odd}
\vspace*{-0.5cm}
\end{figure}

A delocalized state at the center of each Landau level is doubly degenerate when
the valleys are decoupled. A weak valley mixing leads to a small splitting of
the delocalized state \textit{within a single broadened Landau level}. The new
even plateau appears between the two odd ones when the chemical potential lies
between the two split delocalized states (see Fig.\ \ref{fig:odd}). The
longitudinal conductivity $\sigma_{xx}$ has two separated peaks $2\times
g_U^*\simeq e^2/h$ in this case (here the factor $2$ is due to the spin
degeneracy). It is worth mentioning a similarity of the splitting of the
anomalous QHE and the splitting of delocalized states by spin-orbit (spin-flip)
scattering in a spin-degenerate ordinary QHE \cite{spin-orbit}.

\begin{figure}
\centerline{\includegraphics[width=0.9\columnwidth]{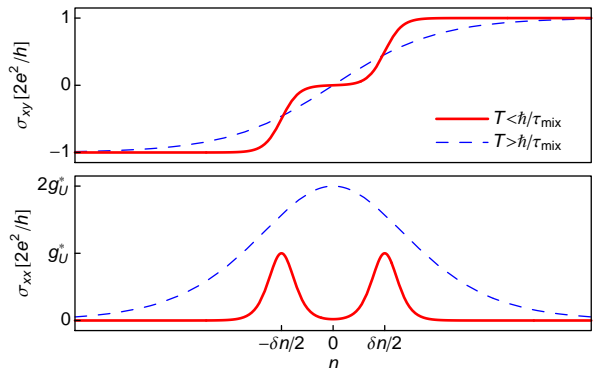}}
\vspace*{-0.2cm}
\caption{Quantum Hall transition at finite temperature. A double step in
$\sigma_{xy}$ and a double peak in $\sigma_{xx}$ (solid lines) require low
temperature, $T \lesssim \hbar/\tau_\text{mix}$. Otherwise a single broadened
quantum Hall transition is seen (dashed lines).}
\label{fig:finiteT}
\vspace*{-0.5cm}
\end{figure}

The flow of $\sigma_{xx}$ and $\sigma_{xy}$ for both cases of decoupled and
mixed valleys is shown in Fig.\ \ref{fig:flow}. For weakly mixed valleys (solid
lines), a crossover occurs between these two flows at the length $l_\text{mix}$.
The even plateaus are much shorter than the odd ones (\ref{odd}) provided the
valley mixing is weak. If the valleys are completely decoupled, the quantum Hall
transition between two successive odd plateaus has a finite width determined by
the temperature-dependent dephasing length $l_\varphi$. The states close to the
center of Landau level are localized at length that diverges as $l_\text{loc}
\sim l(\delta n)^{-\nu}$ where $\delta n$ is the deviation of the filling factor
$n=2\pi l_B^2 n_e$ from the transition point and $\nu \simeq 2.3$ is the
conventional quantum Hall critical index. The width of the transition is then
$\delta n \sim (l/l_\varphi)^{1/\nu}$. If the valley-mixing length
$l_\text{mix}$ is larger than $l_\varphi$, the even plateaus will be totally
smeared --- the splitting between critical states is smaller than the
delocalized energy region around them. The even plateau becomes visible at
sufficiently low $T$, when $l_\varphi$ exceeds $l_\text{mix}$ (see Fig.\
\ref{fig:finiteT}). Therefore, the width of this new plateau is
\begin{equation}
 \delta n_\text{even}
   \sim \delta n(l_\varphi = l_\text{mix})
   \sim (l/l_\text{mix})^{1/\nu}
   \sim (\tau/\tau_\text{mix})^{1/\nu z}.
\end{equation}

For Coulomb impurities, we estimate a typical value of splitting as $\delta
n_\text{even} \sim 0.05$ for the lowest Landau level and $\sim 0.1$ for higher
levels (see Supplementary Materials). In experiment, the temperature should be
low enough in order to resolve the quantum Hall transition splitting, $T
\lesssim \hbar/\tau_\text{mix}$. This implies $T \lesssim 100$\,mK for the
lowest Landau level and $T \lesssim 1$\,K for higher levels. These values are
in reasonable agreement with weak localization measurements in low magnetic
field \cite{Savchenko07}. At higher temperatures, a broadened double step of
Hall conductivity will be seen instead of two split transitions (Fig.\
\ref{fig:finiteT}).

Recent numerical studies \cite{Koshino07} demonstrated the splitting of quantum
Hall transition in graphene with a combination of potential and bond disorder.
At the same time, the model with only potential disorder was found to show only
the odd QHE in Ref.\ \cite{Koshino07}. On the contrary, our consideration yields
the existence of even plateaus in this case but with a narrower plateau at zero
filling factor (see Supplementary Materials). The zeroth plateau arises due to
Landau level mixing which was discarded in Ref.\ \cite{Koshino07}.

Two other mechanisms, apart from intervalley scattering, can establish the even
quantum Hall plateaus, Zeeman splitting and electron-electron interaction.
Zeeman effect is weak in graphene; however, in Ref.\ \cite{Abanin07} the zero
plateau that emerged in high magnetic field was attributed to this mechanism.
An alternative -- Stoner -- mechanism was advocated in Ref.\ \cite{Jiang}.
Indeed, the repulsive interaction between electrons may result in the Stoner
instability \cite{Haldane07, NomuraMcDonald06} giving rise to spontaneous
breaking of spin and/or valley symmetry. Let us note that this instability would
completely split the Landau level leading to the formation of even quantum Hall
plateaus with the width comparable to that of odd plateaus as the magnetic field
or electron mobility is increased \cite{Haldane07}. This can be used to
experimentally distinguish the Stoner splitting from the disorder-induced
splitting analyzed in the present work.

\textbf{\textit{Chiral disorder: ``classical'' QHE.}}---
So far, we have considered the situation of a generic disorder within each
valley. In Ref.\ \onlinecite{OurPRB} it was shown that once the chiral symmetry
$C_0$ is preserved by the disorder (e.g. ripples), the longitudinal
conductivity at zero energy is exactly $4e^2/\pi h$. External magnetic field
also does not violate the chiral symmetry and hence does not change the value
of conductivity \cite{Hikami93}. This leads us to the conclusion that the
quantum Hall transition occurring at zero filling factor is modified by the
presence of $C_0$ symmetry, since $\sigma_{xy} = 4e^2/\pi h$ differs from the
universal value (\ref{sxx}) characteristic for a normal quantum Hall transition,
$\sigma_{xx} \approx 2e^2/h$.

\begin{figure}
\centerline{\includegraphics[width=0.9\columnwidth]{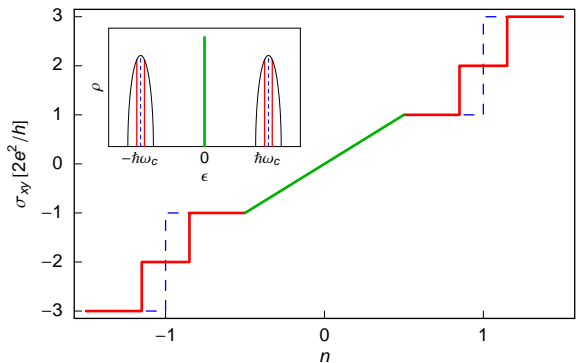}}
\vspace*{-0.2cm}
\caption{``Classical" QHE in graphene with chiral disorder (random vector
potential). Chiral symmetry protects degeneracy of the lowest Landau level
(Inset: delta-function in the density of states). Hall conductivity is a linear
function of carriers concentration while the lowest Landau level is being
filled. In Abelian case (ripples) only odd plateaus appear away from zero energy
(dashed line). Non-abelian gauge disorder (dislocations) split quantum Hall
transitions as shown by solid line.}
\label{fig:chiral}
\vspace*{-0.5cm}
\end{figure}

A general form of chiral disorder in a single valley is a random (Abelian)
vector potential $\mathbf{A}(\mathbf{r})$. The zeroth Landau level remains
exactly degenerate in this situation \cite{AharonovCasher}, as follows from the
Atiyah-Singer theorem \cite{AtyahSinger}. Moreover, one can find explicitly the
wave functions at zero energy (see Supplementary Materials). The exact
degeneracy of the Landau level implies the absence of localization. When the
chemical potential lies at zero energy, the system behaves exactly as if it were
clean. This means that the Hall effect is \emph{classical} rather than quantum
with a linear dependence of Hall conductivity on electron concentration $n_e$
\begin{equation}
 \sigma_{xy}
   = n_e e c/B = n\; 4e^2/h.
 \label{classical}
\end{equation}
This classical dependence holds for filling factor within zeroth Landau level,
$|n| < 1/2$. The longitudinal conductivity remains constant, $\sigma_{xx} = 4
e^2/\pi h$, in this case. The behavior of the Hall conductivity is shown in
Fig.\ \ref{fig:chiral}.

Let us now include a weak valley mixing maintaining the $C_0$ chiral symmetry.
For instance, this is the case when the main disorder due to ripples is
accompanied by rare dislocations. Let us recall that in the case of random
scalar potential, the intervalley mixing leads to the splitting of the quantum
Hall transition into two with a small $\sigma_{xy} = 0$ plateau in between. The
longitudinal conductivity is zero in this case. One could thus expect a similar
behavior for chiral disorder. However, the conductivity at $n=0$ remains
$4e^2/\pi h$ according to the result of Ref.\ \onlinecite{OurPRB} as long as the
chiral symmetry is preserved. This implies no QHE plateau.

How does it happen that the valley mixing does not induce a quantum Hall plateau
around the Dirac point? The answer is the same as for the Abelian random vector
potential discussed above: the zeroth Landau level remains exactly degenerate.
The disorder we consider corresponds to a random \emph{non-Abelian} vector
potential, $\mathbf{A}(\mathbf{r})$, which is a matrix in the valley space. The
degeneracy of the $N=0$ Landau level is a direct corollary of the Atiyah-Singer
theorem \cite{AtyahSinger}. An explicit construction of zero-energy wave
functions \cite{Caux} is given in Supplementary materials. The Hall conductivity
again behaves \emph{classically} within the zeroth Landau level, $\sigma_{xy} =
4ne^2/h$ for $|n| < 1/2$, but the other quantum Hall transitions, away from
$n=0$, split into pairs with narrow plateaus in between (see Fig.\
\ref{fig:chiral}), in the case of weakly mixed valleys.

The observation of a narrow quantum Hall transition in graphene at $n=0$ seems
to indicate that the dominant scattering mechanism is provided by long-range
potential impurities rather than by ripples or dislocations. This is in
agreement with the observed value of the zero-$B$ minimal conductivity at the
Dirac point which is appreciably larger than $4e^2/\pi h$ expected for a random
vector potential. On the other hand, very recent experimental study of quantum
Hall gaps in graphene \cite{Giesbers} revealed that the lowest Landau level is
significantly narrower than other Landau levels. This can be a signature of
preserved chiral symmetry, suggesting that the main scattering mechanism is due
to ripples in the samples studied in Ref.\ \cite{Giesbers}. The quantum Hall
measurement would provide a powerful test of this conjecture.

\textbf{\textit{Double-layer graphene.}}---
Let us turn to the QHE in double-layer graphene. We limit our
consideration to the case of disorder which does not mix the two valleys. The
single-valley Hamiltonian of double-layer graphene reads \cite{McCannFalko}:
\begin{equation}
 H = \frac{1}{2m}[\sigma_x(p_x^2 - p_y^2) + 2 \sigma_y p_x p_y].
\end{equation}

The Landau levels are $\epsilon_N = \hbar \omega_c \sqrt{N(N-1)}$ with the
conventional definition of cyclotron frequency $\omega_c = eB/mc$. The two
lowest levels, $N=0$ and $N=1$, are degenerate. The corresponding wave functions
are spinors in the sublattice space: $(0,\psi_0)^T$ and $(0,\psi_1)^T$,
respectively, where $\psi_N$ is the wave function of $N$-th Landau level in a
normal metal.

\begin{figure}
\centerline{\includegraphics[width=0.9\columnwidth]{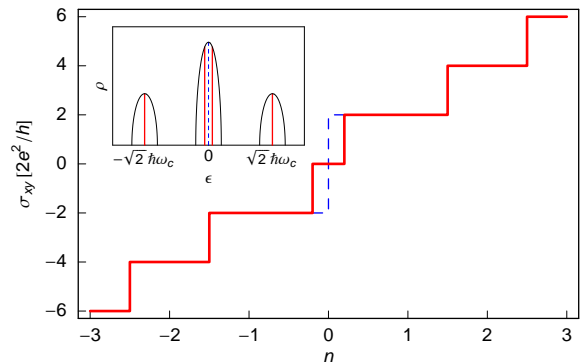}}
\vspace*{-0.2cm}
\caption{QHE in a double-layer graphene with smooth disorder (decoupled
valleys). Degeneracy of the lowest Landau level is twice larger than for other
levels. Double step at zero filling factor (dashed line) is split when the
disorder has finite correlation length $d$. Inset shows the density of states
and positions of delocalized states (solid lines). Two such states within the
lowest Landau level remain degenerate (dashed line) in the limit $d\to \infty$.}
\label{fig:double}
\vspace*{-0.5cm}
\end{figure}

In the presence of a generic disorder within each valley we have the same action
Eq.\ (\ref{action}) but with doubled couplings. The anomalous contribution to
the
topological term gives now $\theta=2\pi$ rather than $\pi$ at zero energy. This
implies complete localization and hence a plateau at $n=0$. However, in
experiments a plateau transition with the double step in $\sigma_{xy}$ at $n =
0$ is observed instead. This can only happen if the disorder does not mix the
two degenerate Landau levels with $N=0$ and $N=1$. The only possible reason of
the lack of mixing is the smoothness of disorder on the scale of magnetic
length $l_B = \sqrt{\hbar c/eB}$. Indeed, the wave functions of the two Landau
levels are orthogonal and concentrated in the area of order $l_B^2$. If the
disorder potential is almost constant in this small region, the corresponding
matrix element is suppressed due to the orthogonality of wave functions. More
specifically, assuming the disorder correlation length $d \gg l_B$, the mixing
rate of the two Landau levels is found as: $\tau_{01}^{-1} \sim \tau^{-1}
(l_B/d)^2$. Comparison of $\tau_{01}$ with the time needed for localization
gives us the width of the zeroth plateau (see Fig.\ \ref{fig:double})
\begin{equation}
 \delta n_0 \sim (l_B/d)^{2/\nu z}.
\end{equation}
To resolve this plateau, one should satisfy an upper bound on temperature,
$T \lesssim \hbar/\tau_{01}$.

It is worth noting that the experimentally measured double step of the Hall
conductivity at $n=0$ can not be automatically explained by the charged
impurities in graphene. A random potential due to charged impurities has no
characteristic length $d$. The only scale associated with such disorder is the
screening length that is of the order of $l_B$ (the stronger is the magnetic
field, the larger is the density of states in Landau level, the more efficient
is the screening) \cite{NomuraMcDonald06}. The experimental observation of the
double step in double-layer graphene thus suggests that an additional scale
exists characterizing the smoothness of disorder. It might be caused by impurity
correlations or, else, by their separation from the graphene layer.

So far, we have considered the QHE in a single valley of a
double-layer sample. If we include an intervalley scattering in our model, than
the $4e^2/h$ quantum Hall steps will further split, similarly to single-layers
studied above. As a result, the conventional QHE with $2e^2/h$
steps will be fully restored.

\textbf{\textit{Summary.}}---
In this paper we have developed the theory of integer QHEs in graphene. The
Landau level structure by itself is not sufficient to determine the form of the
QHE. Anomalous QHEs in graphene are due to special character (symmetry) of
disorder. In particular: (i) a smooth random (scalar) potential which does not
couple the valleys gives rise to the odd QHE, Eq.\ (\ref{odd}) and dashed line
in Fig.\ \ref{fig:odd}; (ii) the valley mixing splits the odd quantum Hall
transitions and restores the ordinary Hall quantization, Eq.\ (\ref{ordinary})
and solid line in Fig.\ \ref{fig:odd}. For weakly mixed valleys the crossover
from the odd to ordinary QHEs occurs at parametrically low but still accessible
temperatures, $T \lesssim \tau_\text{mix}^{-1} \sim 100$\,mK; (iii) ripples or
dislocations (random vector potential preserving the chiral symmetry) lead to a
``classical'' QHE, Eq.\ (\ref{classical}) and Fig.\ \ref{fig:chiral}, around the
half filling; (iv) in double-layers, a double-step QHE transition at $n=0$
arises
for disorder smooth on the scale of $l_B$. Experiments on QHE in graphene thus
provide information about the nature of disorder.

\textbf{\textit{Acknowledgements.}}---
We are grateful to K.S.~Novoselov, L.S.~Levitov, D.G.~Polyakov, and
A.K.~Savchenko for valuable discussions. The work was supported by the Center
for Functional Nanostructures of the Deutsche Forschungsgemeinschaft. The work
of I.V.G. was supported by the EUROHORCS/ESF EURYI Awards scheme.

\clearpage

\onecolumngrid

\centerline{\bfseries\large Theory of Anomalous Quantum Hall Effects in
Graphene}
\vspace{6pt}
\centerline{P.~M.~Ostrovsky, I.~V.~Gornyi, and A.~D.~Mirlin}
\vspace{6pt}
\centerline{\large\itshape Supplementary Materials}
\vspace{18pt}

\twocolumngrid

\textbf{\textit{Disordered graphene in strong magnetic field}}
\nopagebreak
\vspace{6pt}

Here we present the calculation of the averaged Green function and of the
density of states in disordered graphene in the presence of a strong external
magnetic field. We will use the results of this calculation below for the
derivation of the non-linear sigma model. We assume $\hbar = 1$ from now on. 

\vspace{6pt}
\underline{Self-consistent Born approximation}
\nopagebreak
\vspace{6pt}

Let us start with the self-consistent Born approximation (SCBA) approach. We
assume single layer graphene with Gaussian $\delta$-correlated disorder and
consider first the simplest case of potential disorder characterized by a
dimensionless coupling constant $\alpha$ (it corresponds to $\alpha_0$ in Ref.\
\cite{OurPRB}). This type of disorder does not produce any valley mixing, so
that we can use the single-valley Hamiltonian. Intervalley scattering processes
will be included later.

For the single-valley case, the Green function is a $2 \times 2$ matrix in the
sublattice space. In the presence of magnetic field, the disorder-induced
self-energy matrix has two distinct components, $\Sigma_{1,2}$, yielding the
Green function,
\begin{equation*}
 G(\epsilon)
  = \begin{pmatrix}
      \epsilon_1 & v_0 \hat \pi_- \\
      v_0 \hat \pi_+ & \epsilon_2
    \end{pmatrix}^{-1}\!\!,
 \quad
 \hat \pi_\pm
  =  p_x \pm i p_y + \frac{e}{c} (A_x \pm i A_y),
\end{equation*}
where $\epsilon_{1,2} = \epsilon - \Sigma_{1,2}$.

The calculation of inverse matrix is straightforward in the basis of Landau
levels. Owing to the fact that the disorder is $\delta$-correlated, the SCBA
equation involves only the Green function at coincident points. The latter is
independent of a particular gauge and reads
\begin{equation}
 G(\epsilon; \mathbf{r}, \mathbf{r})
  = \frac{\omega_c^2}{4 \pi v_0^2} \begin{pmatrix}
      \epsilon_2 \sum\limits_{N'=1}^\infty & 0\\
      0 & \epsilon_1 \sum\limits_{N'=0}^\infty
    \end{pmatrix}
    \frac{1}{\epsilon_1\epsilon_2 - \omega_c^2 N'}.
\end{equation}
The matrix SCBA equation $\Sigma(\epsilon) = 2 \pi v_0^2 \alpha G(\epsilon;
\mathbf{r}, \mathbf{r})$ determines two self energies $\Sigma_{1,2}$,
\begin{equation}
 \begin{Bmatrix}
  \Sigma_1\\
  \Sigma_2
 \end{Bmatrix}
  = \frac{\alpha \omega_c^2}{2} \begin{Bmatrix}
      \epsilon_2 \sum\limits_{N'=1}^\infty\\
      \epsilon_1 \sum\limits_{N'=0}^\infty
    \end{Bmatrix}
    \frac{1}{\epsilon_1 \epsilon_2 - \omega_c^2 N'}.
 \label{SCBA}
\end{equation}
This equation was analyzed numerically in Ref.\ \cite{ZhengAndo02}. In the
absence of magnetic field, the sum over Landau levels replaces by the integral
and the well-known graphene SCBA equation is reproduced \cite{ShonAndo98}.

We are interested in the case of strong magnetic field, when Landau levels are
well separated. Let us focus on a particular $N$th level. Although SCBA gives
the exact shape of the density of states only for $N \gg 1$, it yields a
parametrically correct estimate for the height and width of the Landau level
peak for all $N$.

Consider first the case $N \neq 0$. Then the largest term in the sums in Eq.\
(\ref{SCBA}) is the one with $N' = N$. We estimate the sum of all other terms by
replacing it with the corresponding integral. (It is worth noting that, contrary
to the case of a normal metal, the contribution of far Landau levels with $N'
\neq N$ can not be neglected. In graphene, the density of states grows linearly
with energy; in strong magnetic field, this leads to Landau level separation
decreasing as $\omega_c/\sqrt{N'}$. As a result, the contribution from high
Landau levels to the self energy should be retained.) The difference between
$\Sigma_1$ and $\Sigma_2$, which originates from the term with $N' = 0$ in Eq.\
(\ref{SCBA}), is immaterial for $N \neq 0$; we will use a unified notation
$\Sigma$ for them. We further simplify the equation by employing the inequality
$|\epsilon - \epsilon_N|, |\Sigma| \ll \epsilon_N$ and obtain
\begin{equation}
 \Sigma
  = \frac{\alpha \omega_c^2}{4 (\epsilon - \epsilon_N - \Sigma)}
    - \alpha (\epsilon - \Sigma) \ln \frac{\Delta}{\epsilon_N}.
 \label{SCBANneq0}
\end{equation}
The logarithmic divergence is cut by the graphene band width $\Delta$.

The effect of magnetic field is encoded in the $\omega_c^2$ term in Eq.\
(\ref{SCBANneq0}). If this term were absent, the result would reproduce the
well-known disorder-driven renormalization of the energy \cite{OurPRB}
\begin{equation}
 \tilde \epsilon
   = \epsilon - \mathop{\text{Re}} \Sigma_0
   = \epsilon/Z,
 \qquad
 Z
   = 1 - \alpha\ln(\Delta/\epsilon_N).
 \label{Z}
\end{equation}
It is instructive to express the solution of the full equation (\ref{SCBANneq0})
in terms of this renormalized energy $\tilde\epsilon$
\begin{equation}
 \epsilon - \Sigma
   = \frac{\tilde\epsilon + \epsilon_N}{2}
     \pm i\sqrt{\tilde \gamma^2 - \left(
       \frac{\tilde\epsilon - \epsilon_N}{2}
     \right)^2}.
\end{equation}
The appeared parameter $\tilde\gamma$ determines the imaginary part of the self
energy in the center of Landau level
\begin{equation}
 \tilde\gamma
   = \frac{\omega_c \sqrt{\alpha}}{2 \sqrt{1 - \alpha\ln(\Delta/\epsilon_N)}}.
 \label{gamma}
\end{equation}

The density of states within each Landau level has a standard form of
semi-circle. The identity $\rho(\epsilon) = -\pi^{-1} \mathop{\text{Im}}
\mathop{\text{tr}} G^R(\epsilon; \mathbf{r}, \mathbf{r})$ together with the
self-consistency equation yields
\begin{multline}
 \rho(\epsilon)
   = \mathop{\text{Im}}\frac{\Sigma_1 + \Sigma_2}{2 \pi^2 v_0^2 \alpha}
   = \frac{\sqrt{4 \tilde\gamma^2 - (\tilde\epsilon - \epsilon_N)^2}}
          {2\pi^2 v_0^2 \alpha} \\
   = \frac{\sqrt{4 \gamma^2 - (\epsilon - \epsilon_N Z)^2}}
          {4 \pi^2 l_B^2 \gamma^2}.
 \label{rho}
\end{multline}
The last expression contains two parameters: the renormalization factor 
$Z$ determines the rescaling of
Landau levels according to Eq.\ (\ref{Z}) and the electrons scattering rate 
$\gamma = \tilde\gamma Z$ gives the Landau level width.

The result (\ref{rho}) together with Eq.\ (\ref{gamma}) provides the following
criterion for the separation of Landau levels within SCBA: the $N$th level
becomes isolated when $\omega_c Z/\sqrt{N} \gtrsim \gamma$. The stronger is the
magnetic field the larger is the number of isolated Landau levels in the
vicinity of the Dirac point.

The solution of the SCBA equation is qualitatively different for the lowest
Landau level, $N = 0$. The distinction between $\Sigma_1$ and $\Sigma_2$ is now
crucial, since the term with $N' = 0$ is absent in the sum (\ref{SCBA}) for
$\Sigma_1$. Replacing the sums over non-zero levels with the integrals, we get
\begin{equation}
\begin{aligned}
 \Sigma_1
   &= - \alpha (\epsilon - \Sigma_2) \ln (\Delta/\omega_c), \\
 \Sigma_2
   &= \frac{\alpha \omega_c^2}{2 (\epsilon - \Sigma_2)}
     - \alpha (\epsilon - \Sigma_1) \ln (\Delta/\omega_c).
\end{aligned}
\label{SCBAN0}
\end{equation}
We express the solution of these equations in terms of renormalized energy
$\tilde\epsilon$ according to Eq.\ (\ref{Z})
\begin{equation}
\begin{aligned}
 \epsilon - \Sigma_2
   &= \frac{\tilde\epsilon}{2}
      \pm i \sqrt{\tilde\gamma_2^2
       - \frac{\tilde\epsilon^2}{4}}, \\
 \Sigma_1
   &= - \alpha \ln(\Delta/\omega_c) \left[
        \frac{\tilde\epsilon}{2}
        \pm i \sqrt{\tilde\gamma_2^2 - \frac{\tilde\epsilon^2}{4}}
      \right].
\end{aligned}
\end{equation}
We denote the imaginary parts of $\Sigma_{1,2}$ at the center of  Landau level
by $\tilde\gamma_{1,2}$
\begin{equation}
 \begin{Bmatrix}
   \tilde\gamma_1 \\
   \tilde\gamma_2
 \end{Bmatrix}
  = \begin{Bmatrix}
      \alpha \ln(\Delta/\omega_c) \\
      1
    \end{Bmatrix} \frac{\omega_c \sqrt{\alpha}}
                       {\sqrt{2[1 - \alpha^2\ln^2(\Delta/\omega_c)]}}.
 \label{g12}
\end{equation}
The electron scattering rates for the two sublattices are given by $\gamma_{1,2}
= Z \tilde\gamma_{1,2}$ with $Z$ from Eq.\ (\ref{Z}). The rate $\gamma_2$ has
the same order of magnitude as for a non-zero Landau level (\ref{gamma}) while
$\gamma_1$ is somewhat smaller. This is a manifestation of the fact that the
lowest Landau level wave function has its support in the sublattice $B$. With
the opposite orientation of magnetic field, the wave function will be in $A$
sublattice and $\gamma_1 > \gamma_2$. In the second valley the situation is
reversed.

Calculating the density of states at the lowest Landau level separately
for two sublattices, we find
\begin{equation}
 \begin{Bmatrix}
   \rho_1 \\
   \rho_2
 \end{Bmatrix}
  = \begin{Bmatrix}
      \alpha \ln(\Delta/\omega_c) \\
      1
    \end{Bmatrix} \frac{\sqrt{4\gamma^2 - \epsilon^2}}
                       {4\pi^2 v_0^2 \alpha Z}.
 \label{rhoSCBAN0}
\end{equation}
For both sublattices, the width of the zeroth Landau level is $\gamma =
\tilde\gamma_2 Z$, which determines the width of the total density of states
$\rho=\rho_1+\rho_2$.

\vspace{6pt}
\underline{Ballistic renormalization group}
\nopagebreak
\vspace{6pt}

As we discuss in the previous section, high Landau levels produce logarithmic
corrections to the low-energy properties of the system. The SCBA takes these
corrections into account only partially. The systematic way for summing up such
logarithms is the renormalization group (RG) formalism
\cite{DisorderedDiracFermions, Bocquet00Altland02, Aleiner06, OurPRB}. Below we
develop this approach for the case of strong magnetic field. As we demonstrate,
the results are qualitatively similar to the SCBA, but differ quantitatively.

In the simplest case of diagonal Gaussian disorder $\alpha$, the starting point
for the renormalization group is the fermionic action
\begin{equation}
 S[\psi]
  = \int d^2r \left[
      -i \bar\psi (\epsilon + i 0 \Lambda - H) \psi
      +\pi v_0^2 \alpha (\bar\psi \psi)^2
    \right].
 \label{fermion}
\end{equation}
Here $H$ is the single-valley Dirac Hamiltonian. The field $\psi$ is an
8-supervector with the structure in the inner AB space (sublattices) of the
Hamiltonian $H$, retarded-advanced (RA) space, and Bose-Fermi (BF) superspace
\cite{Aleiner06,OurPRL}. We use standard notation $\Lambda =
\mathop{\text{diag}}\{1,-1\}_{RA}$. The doubling of variables in the RA space
is needed for the calculation of averages involving both retarded and advanced
Green functions, e.g.\ conductivity. In the ballistic regime that we consider
here, the distinction between retarded and advanced propagator is immaterial.

The renormalization procedure eliminates fast degrees of freedom thus reducing
the cutoff energy $\Delta \to \Delta/L$. The parameters of the action
(\ref{fermion}) are then rescaled according to \cite{OurPRB}
\begin{equation}
 \frac{d \alpha}{d \ln L}
  = 2 \alpha^2,
 \qquad
 \frac{d \epsilon}{d \ln L}
  = \alpha \epsilon.
 \label{alphaRG}
\end{equation}
To study the properties of $N \neq 0$ Landau level, we stop the renormalization
at $L = \Delta/\epsilon_N$, when the running cutoff reaches the observation
energy. The new parameters are
\begin{equation}
 \tilde\alpha
  = \frac{\alpha}{Z^2},
 \qquad
 \tilde\epsilon
  = \frac{\epsilon}{Z},
 \qquad
 Z
  = \sqrt{1 - 2\alpha \ln\frac{\Delta}{\epsilon_N}}.
\end{equation}
Now we employ the SCBA equation (\ref{SCBANneq0}) with the renormalized
parameters. The logarithmic term is absent as long as the running cutoff equals
$\epsilon_N$ after renormalization. The SCBA equation involves a single Landau
level and yields the renormalized self energy $\tilde\Sigma(\tilde\epsilon)$. 
In the center of Landau level, the imaginary part of $\tilde\Sigma$ is
\begin{equation}
 \tilde\gamma
  = \omega_c \sqrt{\tilde\alpha}/2.
 \label{gammaRG}
\end{equation}
The energy dependence of $\mathop{\text{Im}}\tilde\Sigma$ gives the renormalized
density of states $\tilde\rho(\tilde\epsilon)$. In order to calculate the
observable density of states, we use the identity $\rho / \tilde\rho = \partial
\tilde\epsilon / \partial \epsilon$ and obtain
\begin{equation}
 \rho(\epsilon)
  = \frac{\sqrt{\alpha\omega_c^2 - (\epsilon - \epsilon_N Z)^2}}
         {2\pi^2 v_0^2 \alpha}.
 \label{rhoNneq0}
\end{equation}
This result has the same form as the result of SCBA (\ref{rho}), but the
parameters $\gamma = \tilde\gamma Z$ and $Z$ are modified.

At the lowest Landau level, we use equations (\ref{SCBAN0}) with renormalized
parameters and omitted logarithmic terms. We calculate two self-energies,
$\tilde\Sigma_{1,2}$; their imaginary parts at $\epsilon = 0$ are
\begin{equation}
 \tilde\gamma_1
  = 0,
 \qquad
 \tilde\gamma_2
  = \omega_c \sqrt{\tilde\alpha/2}.
 \label{g12RG}
\end{equation}
The renormalized density of states is concentrated in the sublattice B,
\begin{equation}
 \tilde\rho_1
  = 0,
 \qquad
 \tilde\rho_2
  = \frac{\sqrt{2 \tilde\alpha \omega_c^2 - \tilde\epsilon^2}}
    {4 \pi^2 v_0^2 \tilde\alpha}.
 \label{trhoN0}
\end{equation}
In order to find the observable densities, we have to modify our RG scheme.
Different values of $\tilde\rho_{1,2}$ call for introducing the two different
energies, $\epsilon_{1,2}$, in two sublattices. The equations for these energies
have the form
\begin{equation}
 \frac{d \epsilon_1}{d \log L}
  = \alpha \epsilon_2.
 \qquad
 \frac{d \epsilon_2}{d \log L}
  = \alpha \epsilon_1.
\end{equation}
The solution reads
\begin{equation}
 \tilde\epsilon_{1,2}
  = \frac{1}{2} \left[
      \epsilon_{1,2} (Z + Z^{-1}) + \epsilon_{2,1} (Z - Z^{-1})
    \right].
 \label{teps12}
\end{equation}
The connection between $\rho$ and $\tilde\rho$ has the form
\begin{equation}
 \rho_\nu
  = \sum\limits_{\mu=1,2}
    \frac{\partial \tilde\epsilon_\mu}{\partial \epsilon_\nu} \tilde\rho_\mu.
 \label{rhotrho}
\end{equation}
Using Eqs.\ (\ref{trhoN0}), (\ref{teps12}), and (\ref{rhotrho}), we obtain the
resulting density of states at the lowest Landau level in two sublattices
\begin{equation}
 \begin{Bmatrix}
   \rho_1 \\
   \rho_2
 \end{Bmatrix}
  = \begin{Bmatrix}
      \alpha \ln(\Delta/\omega_c) \\
      1 - \alpha \ln(\Delta/\omega_c)
    \end{Bmatrix} \frac{\sqrt{2\alpha\omega_c^2 - \epsilon^2}}
                       {4\pi^2 v_0^2 \alpha}.
 \label{rho12SCBA}
\end{equation}
This result is also similar to its SCBA counterpart Eq.\ (\ref{rhoSCBAN0}) with
slightly modified parameters. The width of the Landau level is $\gamma =
\tilde\gamma_2 Z$. The lowest Landau level becomes isolated when $\omega_c$
exceeds $\Delta e^{-1/2\alpha}$.

We can further improve the result by employing the exact density of states at
the lowest Landau level found by Wegner \cite{Wegner}. After eliminating all
non-zero Landau levels with the help of RG, we find ourselves in the situation
when the approach of Ref.\ \cite{Wegner} is directly applicable and yields
\begin{equation}
 \tilde\rho_2
  = \frac{F(\tilde\epsilon/\tilde\gamma_2)}
         {2\pi^2 l_B^2 \tilde\gamma_2},
 \quad
 F(x)
  = \frac{(2/\sqrt{\pi}) e^{x^2}}
         {1 + \left[ (2/\sqrt{\pi}) \int_0^x e^{y^2} dy \right]^2}.
 \label{F}
\end{equation}
By substituting this result into Eq.\ (\ref{rhotrho}), we calculate the
observable density of states 
\begin{equation}
 \begin{Bmatrix}
   \rho_1 \\
   \rho_2
 \end{Bmatrix}
  = \begin{Bmatrix}
      \alpha \ln(\Delta/\omega_c) \\
      1 - \alpha \ln(\Delta/\omega_c)
    \end{Bmatrix} \frac{F(\epsilon/\gamma)}{2\pi^2 l_B^2 \gamma}.
\end{equation}
This improves the result Eq.\ (\ref{rho12SCBA}) by replacing the semi-circle
function $F(x) = \sqrt{1 - x^2/4}$ with the exact lowest Landau level shape Eq.\
(\ref{F}).

\vspace{6pt}
\underline{Charged impurities}
\nopagebreak
\vspace{6pt}

So far, we have considered finite-range disorder. The model of long-range
charged impurities can also be treated in the framework of SCBA once the
screening is taken into account. For low-lying Landau levels, the screening
length is of the order of magnetic length, which is the only scale in magnetic
field \cite{NomuraMcDonald06}. For high Landau levels, the screening occurs at a
scale of the electron wavelength. If the dimensionless parameter characterizing
the interaction strength is small, $r_s = e^2 / v_0 \chi \ll 1$ ($\chi$ is the
dielectric constant), the screening can be controllably treated within the
random phase approximation. In a more realistic situation, $r_s \sim 1$, the
results for charged impurities are valid up to a numerical factor of order unity
in the definition of the effective disorder strength
\begin{equation}
 \alpha
  \sim n_\text{imp} l_B^2 \begin{cases}
      1/N, & N \neq 0, \\
      1,   & N = 0.
    \end{cases}
 \label{alpha}
\end{equation}

The density of states follows from the SCBA equations (\ref{SCBANneq0}) or
(\ref{SCBAN0}) with $\alpha$ from Eq.\ (\ref{alpha}) and without the
logarithmic terms [i.e., $\ln(\Delta/\epsilon)$ is replaced by a number of
order unity]. The result has the form of Eq.\ (\ref{rhoNneq0}) with $Z = 1$. The
absence of the logarithmic terms is due to the suppression of scattering off
Coulomb impurities at large momentum transfer (that is, transitions involving
far Landau levels are ineffective). The lack of hard scattering also leaves no
room for ballistic renormalization.

\vspace{12pt}
\textbf{\textit{Derivation of the sigma model}}
\nopagebreak

\vspace{6pt}
\underline{Sigma model in a single valley}
\nopagebreak
\vspace{6pt}

Non-linear sigma model is an effective low-energy theory describing soft modes
of the system, diffusons and Cooperons \cite{Efetov}. In the absence of valley
mixing, the sigma model for graphene in zero magnetic field was derived in Ref.\
\cite{OurPRL}. Here we generalize this derivation, allowing for the magnetic
field within a single valley. Then we will also include the intervalley
scattering.

We start the derivation from the fermionic action (\ref{fermion}). The RA
structure of the fields will play a crucial role in the sigma model. Our
calculation is based on the SCBA approach outlined above. The more rigorous RG
calculation can also be used (as in the zero-B case \cite{Aleiner06})
as a basis for the sigma model, leading to the same form of the theory.

The $(\bar\psi \psi)^2$ term in Eq.\ (\ref{fermion}) is decoupled with the help
of an auxiliary $8 \times 8$ supermatrix field $R$. Subsequent Gaussian
integration over $\psi$ yields an effective action in terms of $R$,
\begin{equation}
 S[R]
  = \frac{\mathop{\text{Str}} R^2}{4\pi v_0^2 \alpha}
    + \mathop{\text{Str}} \ln \left[
      \epsilon - H - R
    \right].
\label{log_action_R}
\end{equation}

The soft modes of the system, that sigma model deals with, describe the
fluctuation near the saddle point of $S[R]$. This saddle point is determined by
the self-consistency equations (\ref{SCBA}) with the self energy $\Sigma$
replaced by the matrix $R$. We separate the real and imaginary parts of the
self energy
\begin{equation}
 R
  = \mathop{\text{Re}} \Sigma + i \tilde\Gamma \Lambda,
 \label{R}
\end{equation}
where $\tilde\Gamma$ is the matrix of renormalized scattering rates,
$\tilde\Gamma = \mathop{\text{diag}}\{\tilde\gamma_1, \tilde\gamma_2\}_{AB}$,
given by Eqs.\ (\ref{gamma}) or (\ref{g12}). A whole saddle manifold can be
generated from the solution (\ref{R}) by a uniform rotation $T$ that commute
with the Hamiltonian $H$. As a result, the matrix $\Lambda$ in the imaginary
part of Eq.\ (\ref{R}) replaces with $Q = T^{-1} \Lambda T$. The $4 \times 4$
matrix $Q$ operates in RA and BF spaces and obeys the constraints
$\mathop{\text{str}}Q = 0$ and $Q^2 = 1$. We rewrite the action
(\ref{log_action_R}) in terms of $Q$ omitting the first term which produces an
unphysical constant,
\begin{equation}
 S[Q]
  = \mathop{\text{Str}} \ln \left[
      \tilde\epsilon - H + i \tilde\Gamma Q
    \right],
 \label{log_action}
\end{equation}
The real part of the self energy is included in $\tilde\epsilon$ which becomes
an AB matrix. Effective low-energy action (sigma model) is a result of the
gradient expansion of Eq.\ (\ref{log_action}). This expansion is a non-trivial
procedure in view of the topology of the saddle manifold \cite{Pruisken84}.
Furthermore, the Dirac nature of electrons in graphene gives rise to extra
anomalous contributions to the sigma-model action \cite{OurPRL}.

The approach of Ref.\ \cite{OurPRL} is directly applicable to the derivation of
the sigma model in magnetic field. The key feature of this approach is a special
form of boundary conditions involving the mass term, $m \sigma_3$, in the
Hamiltonian. Assuming the mass is zero in the bulk of the sample and gradually
increases up to some large value $M$ near the boundary, we get the sigma-model
action \cite{OurPRL}
\begin{equation}
  S[Q]
  = \frac{1}{4} \mathop{\mathrm{Str}} \left[
       -\frac{g_{xx}}{2} (\nabla Q)^2
       +\frac{\theta}{2 \pi}  Q \nabla_x Q \nabla_y Q
    \right],
\end{equation}
with the topological angle
\begin{equation}
 \theta
  = 2\pi[g^I_{xy}(0) + g^{II}_{xy}(0) - g^{II}_{xy}(M)].
 \label{theta}
\end{equation}
The parameters of the model are determined by the standard Kubo expressions
\begin{align}
 g_{xx}
  &= -\frac{1}{2} \mathop{\text{Tr}} \left[
      j_x (G^R - G^A) j_x (G^R - G^A)
    \right],\\
 g^I_{xy}
  &= -\frac{1}{2} \mathop{\text{Tr}} \left[
      j_x (G^R - G^A) j_y (G^R + G^A)
    \right],\\
 g^{II}_{xy}
  &= \frac{i e}{2} \mathop{\text{Tr}} \left[
      (x j_y - y j_x) (G^R - G^A)
    \right]. \label{gIIxy}
\end{align}
Trace in the last equation is divergent and requires a regularization. This
happens because $g^{II}_{xy}$ accounts for the contribution of edge modes to
Hall conductivity. That is why we have to specify boundary conditions in order
to find $g^{II}_{xy}$.

The dependence of $g^{II}_{xy}$ on boundary conditions shows that the very
notion of the single-valley Hall conductivity can not be properly defined. The
observable Hall conductivity 
\begin{equation}
 g_{xy}
  = g^I_{xy} + \frac{1}{2}\Bigl(
      g^{II}_{xy} - \left. g^{II}_{xy}\right|_{B \to -B}
    \Bigr)
\end{equation}
always includes contributions from both, mutually time-reversed, valleys
implying
a cancellation of divergences in Eq.\ (\ref{gIIxy}). Considering the Hall
conductivity per valley, one usually means a half of the total observable Hall
conductivity. This corresponds to a certain regularization requiring
$g^{II}_{xy} = 0$ at the Dirac point.

At the same time, the value of $\theta$ in the sigma-model action is
well-defined (modulus $2\pi$) even within a single valley as long as $\theta$
contains a difference of two $g^{II}_{xy}$ quantities (\ref{theta}). At the
boundary, the introduced mass $M$ is large, so we can neglect energy and
magnetic field there and obtain $g^{II}_{xy}(M) - g^{II}_{xy}(0) = (1/2)
\mathop{\text{sign}} M$. This provides the anomalous topological term in the
sigma model Eq.\ (\ref{action}) with $\theta=g_{xy}+1/2 \mathop{\text{sign}} M$.
The sign of the anomalous term ($\mathop{\text{sign}} M$ here)  is immaterial as
it only changes the action by an integer multiple of $2\pi i$.

It is worth emphasizing that the localization or criticality is the property of
the bulk theory and does not depend on the boundary condition. Nevertheless,
similarly to the ordinary QHE, introducing the boundary turns out to be a
convenient way of deriving the field theory, since the action contains a
topological term. The resulting theory however does not depend on whether a
system with boundary or without it (say, on a sphere) is considered and on the
way the boundary is implemented. Indeed, the final form of the topological term
in Eq.\ (\ref{action}) is represented as a 2D integral over the bulk. Thus the
boundary only facilitates revealing and exploring the intrinsic topological
properties of Dirac fermions in the bulk of graphene.

An alternative derivation of the sigma model for Dirac fermions employs
non-Abelian bosonization \cite{Bocquet00Altland02} that does not require an
introduction of boundary conditions. In bosonic language, disorder leads to
constraint on the boson field reducing the chiral gauge symmetry group down to
sigma-model manifold. The Wess-Zumino term in the bosonized action transforms
into the anomalous topological term of the sigma model. This method was used in
Ref.\ \cite{Altland06} for graphene with mixed valleys.

\vspace{6pt}
\underline{Intervalley scattering}
\nopagebreak
\vspace{6pt}

Let us now add an intervalley scattering term to the fermionic action
(\ref{fermion}). The intervalley scattering due to time-reversal invariant
disorder is described by two coupling constants, $\beta_{\perp}$ and $\beta_z$
(see Ref.\ \cite{OurPRB} for details),
\begin{multline}
 S_\text{mix}
  = 2 \pi v_0^2 \\
    \times \mathop{\text{Str}}\Bigl\{ \beta_\perp \left[
      (\psi \bar\psi)_{AK} (\psi \bar\psi)_{AK'}
      +(\psi \bar\psi)_{BK} (\psi \bar\psi)_{BK'}
    \right] \\
    + \beta_z \left[
      (\psi \bar\psi)_{AK} (\psi \bar\psi)_{BK'}
      +(\psi \bar\psi)_{BK} (\psi \bar\psi)_{AK'}
    \right] \Bigr\}.
 \label{Smix}
\end{multline}
We will treat this term perturbatively within the SCBA scheme. This is
equivalent to replacing a pair of $\psi$ fields with the corresponding Green
function which, on the saddle-point level, is equal to the matrix $Q$:
\begin{equation}
 (\psi \bar\psi)_{K,K'}
  \mapsto \frac{\tilde\Gamma_{K,K'} Q_{K,K'}}{2\pi v_0^2 \alpha}.
 \label{saddlemap}
\end{equation}
The imaginary part of self energy is different in two valleys, $\tilde\Gamma_K =
\mathop{\text{diag}}\{\tilde\gamma_1, \tilde\gamma_2\}_{AB}$, $\tilde\Gamma_{K'}
= \mathop{\text{diag}}\{\tilde\gamma_2, \tilde\gamma_1\}_{AB}$. After the
substitution (\ref{saddlemap}), the valley-mixing action acquires the form of
Eq.\ (\ref{S12}). We calculate the mean free time from the width of Landau
level, $\tau = 1/(4\gamma)$, and obtain in the level's center
\begin{equation}
 \frac{\tau_\text{mix}}{\tau}
  = \frac{2 \omega_c \alpha^2}
         {2\pi \beta_\perp \tilde\gamma_1 \tilde\gamma_2
          +\pi \beta_z (\tilde\gamma_1^2 + \tilde\gamma_2^2)}.
 \label{taumix}
\end{equation}

The form of the $S_\text{mix}$ term is universal and does not rely on the
particular disorder model. At the same time, the mixing rate $\tau_\text{mix}$
is determined by microscopic non-universal mechanisms and depends on the 
disorder type. A potential disorder provides only the intervalley coupling
$\beta_\perp$. Using the SCBA results (\ref{gamma}) and (\ref{g12}), we find
\begin{equation}
 \frac{\tau_\text{mix}}{\tau}
  = \begin{cases}
      \dfrac{4\alpha[1 - \alpha \ln(\Delta/\epsilon_N)]}{\pi\beta_\perp},
       & N \neq 0, \\[0.3cm]
      \dfrac{2[1 - \alpha^2 \ln^2(\Delta/\omega_c)]}
            {\pi\beta_\perp\ln(\Delta/\omega_c)},      
       & N = 0.
    \end{cases}
 \label{taumixSCBA}
\end{equation}

In order to apply the ballistic RG approach, we first renormalize the action
(\ref{fermion}) including the mixing term (\ref{Smix}). Assuming the inequality
$\alpha \gg \beta_{\perp,z}$, we employ the simplified version of RG equations
\cite{Aleiner06, OurPRB}
\begin{equation}
 \frac{d\beta_\perp}{d\ln L}
  = 4 \alpha \beta_z,
 \qquad
 \frac{d\beta_z}{d\ln L}
  = -2 \alpha \beta_z + 2 \alpha \beta_\perp
 \label{RGbeta}
\end{equation}
in addition to Eq.\ (\ref{alphaRG}). The renormalized couplings are then
substituted into Eq.\ (\ref{taumix}) with the parameters $\tilde\gamma_{1,2}$
given by Eqs.\ (\ref{gammaRG}) and (\ref{g12RG}). In terms of bare couplings
$\alpha$ and $\beta_\perp$ (potential disorder), the mixing time is
\begin{equation}
 \frac{\tau_\text{mix}}{\tau}
  = \begin{cases}
      \dfrac{4\alpha}{\pi\beta_\perp},
       & N \neq 0, \\[0.3cm]
      \dfrac{2}{\pi\beta_\perp\ln\frac{\Delta}{\omega_c}[
        1 - 2\alpha\ln\frac{\Delta}{\omega_c}
        +\frac{4\alpha^2}{3} \ln^2\frac{\Delta}{\omega_c}
      ]},     
       & N = 0.
    \end{cases}
 \label{taumixRG}
\end{equation}
For the lowest Landau level, the RG rate $\tilde\gamma_1$ appears to be zero
\cite{Koshino07} since the wave function resides solely in the sublattice B.
The valley mixing occurs only due to $\beta_z$ disorder. For potential
impurities, this coupling has been absent in the ultraviolet limit but is
generated by the RG flow (\ref{RGbeta}).

The SCBA and RG results (\ref{taumixSCBA}) and (\ref{taumixRG}) coincide up to a
numerical factor of order unity once the Landau levels are well separated, i.e.
in the range of our interest. The criterion of level separation is provided by
RG calculation: $\omega_c > \Delta e^{-1/2\alpha}$.

\vspace{12pt}
\textbf{\textit{Estimation of $\tau_\text{mix}$}}
\nopagebreak
\vspace{6pt}

Now we apply the model of screened Coulomb impurities to find the values of
$\tau$ and $\tau_\text{mix}$. This model is most relevant for graphene
experiments because it conforms with both linear dependence of conductivity on
the concentration of electrons \cite{NomuraMcDonald06} and with minimal
conductivity at the Dirac point \cite{OurEPJ}.

The Born parameter of screened Coulomb impurities is given by Eq.\
(\ref{alpha}). Intervalley scattering involves large momentum transfer. Thus
we neglect screening and estimate $\beta_\perp \sim n_\text{imp} a^2$ where $a$
is the lattice constant. The valley mixing rate follows from any of Eqs.\
(\ref{taumixSCBA}) or (\ref{taumixRG}) with logarithmic factors replaced by some
numbers of order unity. 
\begin{equation}
 \frac{\tau_\text{mix}}{\tau}
  \sim \begin{cases}
       l_B^2/a^2 N, & N \neq 0, \\
       (n_\text{imp} a^2)^{-1}, & N = 0.
     \end{cases}
\end{equation}
Taking a typical magnetic field value of 20\,T, we get a 10\% splitting of
quantum Hall transitions for non-zero Landau levels. Another splitting of order
5\% appears at zero Landau level if we estimate $n_\text{imp} \sim 4 \times
10^{11}\,\text{cm}^{-2}$ from mobility measurements away from the Dirac point
\cite{Novoselov05}.

To observe the quantum Hall transition splitting, the temperature should be
small enough. Namely, one should have $T \lesssim 1/\tau_\text{mix}$. For
non-zero Landau level at 20\,T this gives an upper bound of $\sim 1$\,K. The
splitting of zero level becomes visible at smaller temperatures $\sim 100$\,mK.

\vspace{12pt}
\textbf{\textit{Wave functions in chiral disorder}}
\nopagebreak
\vspace{6pt}

\underline{Abelian case}
\nopagebreak
\vspace{6pt}

Abelian chiral disorder has a form of random vector potential. After a proper
gauge transformation, any two-dimensional vector potential can be expressed as
a curl of a scalar field $\phi(\mathbf{r})$
\begin{equation}
 A_x
  = -\nabla_y \phi,
 \qquad
 A_y
  = \nabla_x \phi.
\end{equation}
This field $\phi$ is uniquely determined by the magnetic field $B(\mathbf{r})$
penetrating the system, $\nabla^2 \phi = - B$. Assume that the uniform part of
magnetic field $B_0$, the one that establishes Landau levels, is pointing up,
$B_0 > 0$. Then the function $\phi$ grows at infinity as $\phi \sim B_0 r^2$ and
all the zero-energy wave functions lie entirely in the sublattice B. A possible
set of such functions (up to a normalization factor) is
\begin{equation}
 \Psi_m^B(x,y)
  = (x - iy)^m \exp(-e\phi/c).
\end{equation}

\vspace{6pt}
\underline{Non-Abelian case}
\nopagebreak
\vspace{6pt}

Non-Abelian vector potential has a matrix structure in the valleys space. An
explicit construction of zero-energy wave functions is almost the same as above
\cite{Caux}: express the vector potential in the form
\begin{equation}
 A_+
  = \frac{i c}{e}\, g^{-1} \partial_+ g,
 \qquad
 A_-
  = -\frac{i c}{e}\, g \partial_- g^{-1},
\end{equation} 
where $A_\pm = A_x \pm i A_y$, $\partial_\pm = \nabla_x \pm i \nabla_y$, and $g$
is an appropriate $2 \times 2$ matrix in the valleys space. The wave functions
of the zeroth Landau level again lie in the sublattice B and have the form
\begin{equation}
 \Psi_m^B(x,y)
  = (x - iy)^m g_{1,2}
\end{equation}
with $g_{1,2}$ being any of the two columns of the matrix $g$.

\end{document}